\documentclass[mathleft,fleqn]{an}

\usepackage{graphicx}
\usepackage[varg]{txfonts}
\usepackage{amssymb}

\newcommand{\be}{\begin{equation}}
\newcommand{\ee}{\end{equation}}
\newcommand{\bel}[1]{\be\label{#1}}

\newcommand{\ds}{\displaystyle}
\newcommand{\hsp}{\hspace*{1pt}}
\newcommand{\hspm}{\hspace*{.5pt}}
\newcommand{\ov}[1]{\overline{#1}}

\begin{document}

\Yearpublication{2014}%
\Yearsubmission{2014}%
\Month{0}%
\Volume{999}%
\Issue{0}%
\DOI{asna.201400000}%

\title{Undersaturation of quarks at early stages of relativistic nuclear\\
collisions: the hot glue initial scenario and its observable signatures}

\author{
        H. St\"ocker\inst{1,2,3}\fnmsep\thanks{E-mail:{H.Stoecker@gsi.de}},
        M.~Beitel\inst{2},
        T.S.~Bir\'o\inst{4},
        L.P.~Csernai\hspm\inst{5},
        K.~Gallmeister\hspm\inst{2},
        M.I.~Gorenstein\inst{3,6},\\
        C.~Greiner\hspm\inst{2},
        I.N.~Mishustin\inst{3,7},
        M.~Panero\inst{8},
        S.~Raha\inst{9},
        L.M.~Satarov\hspm\inst{3,7},
        S.~Schramm\hspm\inst{2,3},
        F.~Senzel\hspm\inst{2},\\
        B.~Sinha\inst{10},
        J.~Steinheimer\hspm\inst{3},
        J.~Struckmeier\hspm\inst{1,3},
        V.~Vovchenko\inst{1,3,11},
        Z.~Xu\inst{12},
        K.~Zhou\hspm\inst{2,3},
        P.~Zhuang\inst{12}
 }

\institute{
GSI Helmholtzzentrum f\"ur Schwerionenforschung GmbH, D-64291 Darmstadt, Germany
\and
Institut f\"ur Theoretische Physik, Goethe Universit\"at Frankfurt, D-60438 Frankfurt Main, Germany
\and
Frankfurt Institute for Advanced Studies, D-60438 Frankfurt am Main, Germany
\and
Institute for Particle and Nuclear Physics, Wigner RCP, Budapest, Hungary
\and
Institute for Physics and Technology, University of Bergen,
5007 Bergen, Norway
\and
Bogolyubov Institute for Theoretical Physics, 03680 Kiev, Ukraine
\and
National Research Center ''Kurchatov Institute'', 123182 Moscow, Russia
\and
University of Turin and INFN, Turin, Italy
\and
Bose Institute, Kolkata, India
\and
Variable Energy Cyclotron Centre, Kolkata, India
\and
Taras Shevchenko National University of Kiev, 03022 Kiev, Ukraine
\and
Department of Physics, Tsinghua University, Beijing, China
}

\received{XXXX}
\accepted{XXXX}
\publonline{XXXX}


\abstract{
The early stage of high multiplicity nuclear collisions is represented by a nearly quarkless, hot, deconfined pure gluon plasma. This new scenario should be characterized by a suppression of high $p_T$ photons and dileptons as well as by reduced baryon to meson ratios. We present the numerical results
for central Pb+Pb collisions at the LHC energies by using the ideal Bjorken hydrodynamics with time-dependent quark fugacity. It is shown that about 25\%
of final total entropy is generated during the hydrodynamic evolution of
chemically undersaturated quark-gluon plasma.
  }

\maketitle

\section{Introduction}
The proper understanding of the initial and the early stage of ultra-relativistic pp-, pA- and heavy ion AA- collisions is a topic of great importance for our understanding of hot and dense QCD matter formed in the laboratory and in the early universe. One of the central questions is how the initial highly nonequilibrium system evolves to a state of partial thermodynamic equilibrium at later stages of nuclear collisions. There exist several models which describe the initial state in terms of non-equilibrium parton cascades~(Wang~\& Gyulassy 1991; Xu \& Greiner 2005), minijets~(Eskola~\& Kajantie 1997), color glass condensate~(McLerran~\& Venugopalan 1994), coherent chromofields~(Magas et al. 2001; Mishustin \& Kapusta 2002) etc. It is commonly believed that the strong non-equilibrium effects persist only for
a~short time $\sim 1/Q_s$, where $Q_s\simeq 1-2~\textrm{GeV}$ is the so-called saturation
scale~(Gribov et al. 1983), but at later times the system reaches a state
of a partial thermodynamic equilibrium.

Relatively large gluon-gluon cross sections lead to the idea (van Hove \& Pokorski 1975)
that the gluonic components of colliding nucleons interact more strongly than the quark-antiquark ones. Then the two-step equilibration of the quark-gluon plasma (QGP) was proposed~(Raha 1990; Shuryak 1992; Alam et al. 1994; McLerran \& Venugopalan 1994; Krasnitz \& Venugopalan 2001). In this approach the gluon thermalization takes place at the proper time $\tau_g<1fm/c$ and the (anti)\hspm quark equilibration occurs at $\tau_{\rm th}>\tau_g$.

The surprising result, that only very few soft quarks are present at an early stage of a relativistic collision, was obtained in many transport
calculations (Bir{\'o} et al. 1993; Roy et al. 1997; Elliott \& Rischke 2000; Blaizot et al.~2013; \mbox{Uphoff} et al. 2015). Observable consequences
of the above two-step scenario was considered by several authors, see e.g. (Strickland 1994, K\"ampfer \& Pavlenko 1994, Traxler \& Thoma 1996; Dutta et al. 2002; Gelis et al. 2004; Scardina et al. 2013; Liu et al. 2014; Monnai 2014; Stoecker et al 2015; Vovchenko et al. 2015a). One immediate prediction of such a ''pure glue'' initial scenario is suppressed yields of hard
''thermal'' photons and dileptons\hsp\footnote
{
One should distinguish such particles from photons and Drell-Yan dileptons
produced in inelastic collisions of initial partons.
}.
Such a suppression occurs due to the reduction of the electric charge density as compared to chemically equilibrated quark-gluon plasma (QGP) at early stages of the reaction.

\section{Evolution of undersaturated QGP in nuclear collision}

Below we assume that a thermally (but not necessary chemically) equilibrated QGP
is created initially in a nuclear collision. In this section we use the parameters typical for central
Pb+Pb collisions at the LHC energy $\sqrt{s_{NN}}=2.76~\textrm{TeV}$.
By using the one-dimensional scaling hydrodynamics (Bjorken 1983) we consider the space-time evolution
of QGP produced at the proper time $\tau=\tau_0$. We adopt the equation of state of an ideal gas
of massless gluons, quarks and antiquarks. At zero net baryon density one
may describe deviations from chemical equilibrium for quarks and antiquarks\hsp\footnote
{
In accordance with the two-step approximation (see Sec.~1) we neglect deviations from chemical
equilibrium for gluons, assuming that their fugacity is equal to unity during the whole process
of the QGP evolution at~$\tau\geqslant\tau_0$\hsp.
}
by introducing the quark fugacity $\lambda$. Within these approximations, the following relations for the energy density $\varepsilon$ and pressure $P$ can be written ($\hbar=c=1$)
\bel{eos}
\varepsilon=3P=\sigma T^{\hspm 4},~~~\sigma=\frac{\pi^{\,2}}{30}\left(16+{\lambda}\,\frac{21}{2}\, N_f\right),
 \ee
where $T$ is temperature, $N_f$ is the number of quark flavours (unless stated otherwise, we assume that $N_f=3$). The first and second terms in the last equality describe, respectively, the contributions of gluons and quark-antiquark pairs to the energy density. The parameter $\lambda$ changes from zero for the pure gluonic system to unity for chemically equilibrated~QGP.
\begin{figure}[tp]
\centering
\includegraphics[width=0.9\linewidth]{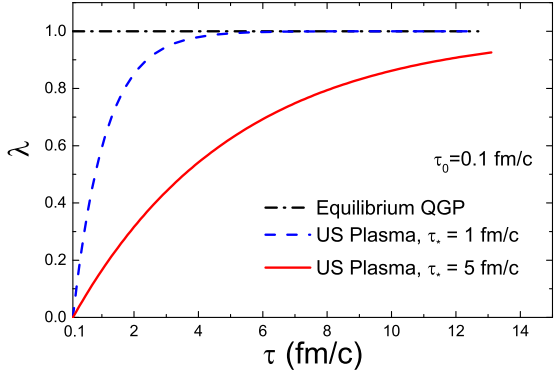}
\caption[]{(Color online)
The quark fugacity $\lambda$ as a function of proper time $\tau$. The solid and dashed lines correspond to chemically undersaturated (US) plasma with parameters $\tau_* = 1~\textrm{fm}/c$ and  $5~\textrm{fm}/c$\hsp, respectively. The dashed-dotted line corresponds to the case of chemical equilibrium ($\lambda=1)$.
}
\label{fig1}
\end{figure}

By using the relation $P=\varepsilon/3$ and neglecting the viscosity effects one can easily get the
analytic solution of the hydrodynamic equation in the Bjorken model
\bel{eps-tau}
\varepsilon~=~\varepsilon(\tau_0)\,\left(\frac{\tau_0}{\tau}\right)^{4/3},
\ee
where the parameter $\tau_0$ corresponds to the initial proper time of the hydrodynamic expansion.
In principle one may determine $\lambda$ and $T$ as functions of $\tau$ by solving numerically
the additional rate equation for quark density evolution (see e.g. Bir\'o et al. 1993; Monnai 2015).
The qualitative analysis can be performed by introducing the
analytic parametrization~(Vovchenko et al. 2015a)
\bel{lambda}
\lambda\hsp (\tau)~=~1~-~\exp\left(\hsp\frac{\tau_0-\tau}{{\tau_*}}\right)\,,
\ee
where $\tau_*$ is the model parameter characterizing the quark chemical equilibration time.
Calculations of different authors lead to different estimates for $\tau_*$\hsp , ranging from
\mbox{$\tau_*\sim 1~\textrm{fm}/c$} (Ruggieri et al. 2015) to $\tau_*\sim 5~\textrm{fm}/c$ (Xu \& Greiner 2005).
One should have in mind that this parameter may depend on the combination of nuclei and the bombarding
energy. We expect that $\tau_*$ will be larger for peripheral events and lighter combinations of nuclei. Figure~\ref{fig1} shows the $\lambda(\tau)$ values for different choices of the parameter
$\tau_*$\hspm .

Introducing the quark chemical potential $\mu=T\ln{\lambda}$ and using thermodynamic relations,
one can write down the following expression for entropy density of the QGP
\bel{ent-den}
s\simeq\frac{32\hsp\pi^{\,2}}{45}\hsp
T^{\hspm 3}\left[1+\lambda\left(0.66-0.16\ln{\lambda}\right)N_f\hsp\right].
\ee
By using (\ref{eos})--(\ref{ent-den}) one can show that $s\tau$ is increasing function of $\tau$, i.e.
$s\tau\geqslant s_0\tau_0$\hsp , where the equality holds only in the equilibrium limit $\lambda=1$.

Within the Bjorken model the total entropy per unit space-time rapidity $\eta=\tanh^{-1}(z/t)$
can be expressed as (\mbox{Satarov} et al. 2007)\hsp\footnote
{
We consider a purely central collision of equal nuclei.
}
\bel{dSdy}
\frac{dS(\tau)}{d \eta} =\pi \hsp R^{\hsp 2}_A \hsp s\hspm (\tau) \hsp \tau,
\ee
where $R_A$ is the geometrical radius of the colliding nuclei.
Therefore, we obtain that the total entropy per unit space-time rapidity is not conserved:
it gradually increases during the system expansion from the pure glue initial state.
Note that this increase occurs within the ideal hydrodynamics, in absence of viscosity
effects. We think that future models for extracting the viscosity values from the observed
data should take into account the suppression of quarks at the
initial state of a nuclear collision.
\begin{figure}[tbh!]
\centering
\includegraphics[width=0.9\linewidth]{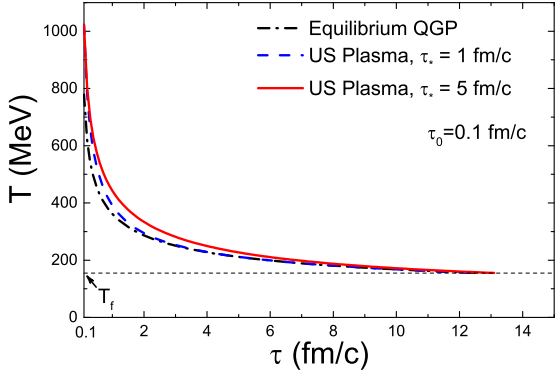}
\caption[]{(Color online)
Temperature of QGP produced in central PbPb collision ($\sqrt{s_{NN}}=2.76~\textrm{TeV}$) as a function of $\tau$. The solid and dashed curves correspond to chemically undersaturated matter assuming the parameters $\tau_* = 1~\textrm{fm}/c$ and  $5~\textrm{fm}/c$\hsp, respectively. The dashed--dotted
line is calculated within the equilibrium scenario.
}
\label{fig2}
\end{figure}

\begin{figure}[tp]
\centering
\includegraphics[width=0.9\linewidth]{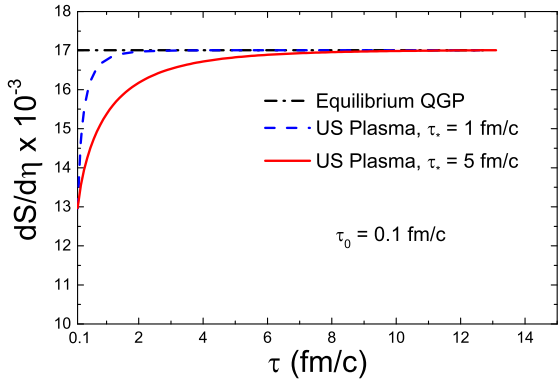}
\caption[]{(Color online)
Same as Fig.~\ref{fig2}, but for the entropy per unit space-time rapidity.
}\label{fig3}
\end{figure}

To fix the initial temperature, we assume that the Bjorken solution is valid
until the freeze-out hypersurface $\tau=\tau_f$. The latter is determined
by the condition $T(\tau_f)=156~\textrm{MeV}$. Such a temperature has been extracted
(Stachel et al. 2014) from the thermal fit of hadron ratios in the considered reaction.
In our calculation we use the approximate relation (Hwa \& Kajantie 1985) between the total
entropy per unit space-time rapidity and the rapidity density of pions
\bel{dsdef}
\frac{dS(\tau_f)}{d\eta}=\nu\,\left.\frac{dN_\pi}{dy}\right|_{y=\eta}\hspace*{-2mm}\simeq 1.7\cdot 10^4,
\ee
where \mbox{$\nu\simeq 6.3$} is the entropy per pion at freeze-out (Vovchenko et al. 2015a) and
\mbox{$dN_\pi/dy\simeq 2700$}~~is the observed yield of pions at midrapidity
(Abb\hspm as et al. 2013)\hspm .

By using this procedure we have calculated the temperature and entropy density of matter in the pure glue initial scenario and compared the results with the chemical equilibrium case. Figure~\ref{fig2} shows
the evolution of temperature for the same values of parameters $\tau_*$ and $\tau_0$ as in
Fig.~\ref{fig1}. One can see that temperature of the US plasma is noticeably higher
than in the equilibrium scenario. This increase is especially visible at $\tau\lesssim\tau_*$\hsp .

Figure~\ref{fig3} shows the results for the total entropy evolution. One can see that
this quantity gradually increases and reaches the freeze-out value (\ref{dsdef}) during the time interval $\Delta\tau\sim\tau_*$\hsp . According to our calculations, the total increase of entropy is not sensitive to $\tau_*$ and equals about 25\%
of the final value.

\section{Dilepton and photon spectra in the pure glue initial scenario}

\begin{figure}[tp]
\centering
\includegraphics[trim=0 1mm 0 0, clip, width=0.9\linewidth]{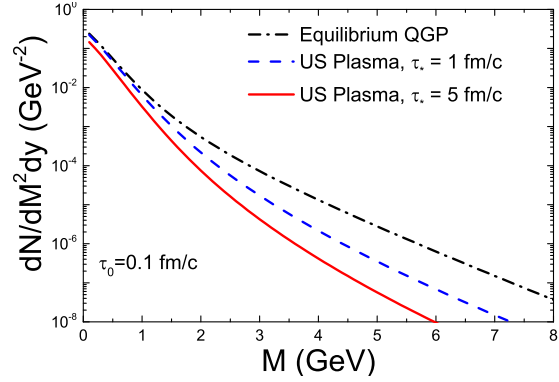}
\caption[]{(Color online)
Mass distribution of thermal dileptons in central PbPb collisions
at $\sqrt{s_{NN}}=2.76~\textrm{TeV}$.
The solid and dashed lines are calculated with parameters $\tau_* = 1~\textrm{fm}/c$,
and $5~\textrm{fm}/c$\hsp , respectively.
The dashed-dotted line corresponds to the chemical equilibrium scenario.
}\label{fig4}
\end{figure}

Hard thermal photons and dileptons are sensitive probes of hot initial stages of high energy nuclear
collisions. We assume that hard dileptons are mostly produced in the deconfined phase via the
\mbox{$q\ov{q}\to l^+l^-$} annihilation processes. Below we again apply the Bjorken hydrodynamics
to describe the evolution of QGP in a heavy-ion collision. As compared to early calculations (Hwa \& Kajantie 1985; K\"ampfer et al. 1990) we include the additional factor $\lambda^2$ which takes into account the reduction of quark and antiquark densities in the chemically nonequilibrium case. We get the following expression for the mass spectrum of $e^+e^-$ pairs:
\begin{eqnarray}
\frac{dN_{e^+e^-}}{dM^{\hsp 2}\,dy}=&&\frac{\alpha^2}{\pi^2}
\sum_{i=u,d,s} q_i^2\,R_A^2\,M\,\times\nonumber\\
&&\times\int_{\tau_0}^{\tau_f}\tau d\tau\,
T(\tau)~K_1[M/T(\tau)]\,\lambda^2(\tau)~,\label{lepton}
\end{eqnarray}
where $\alpha=e^{\hspm 2}$ is the electromagnetic coupling constant, $q_i$ is the
charge of the quark flavour $i$ in units of~$e$ and $K_1(x)$ is the Macdonald function.
As above, current masses of quarks are disregarded for all flavours. Note that different scenarios
considered in Sec. 2 correspond to different choices of $\lambda(\tau)$ and $T(\tau)$.
The results of numerical calculations are shown in Fig.~\ref{fig4}. One can see that
at $M\gtrsim 1~\textrm{GeV}/c^2$ the dilepton spectra are strongly sensitive to chemical nonequilibrium effects. A special investigation shows that such dileptons are mainly produced at hot early stages of the reaction. In the pure glue initial scenario yields of hard dileptons are suppressed as compared to the equilibrium case.

\begin{figure}[tp]
\centering
\includegraphics[width=0.9\linewidth]{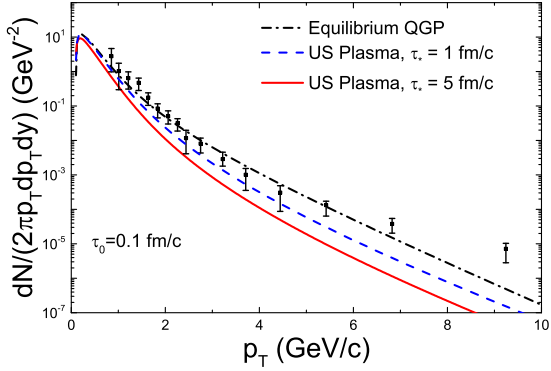}
\caption[]{(Color online)
Spectrum of the thermal photons as a fuction of transverse momentum
in central PbPb collisions at $\sqrt{s_{NN}}=2.76~\textrm{TeV}$.
The solid and dashed lines correspond,
respectively, to $\tau_* = 1~\textrm{fm}/c$ and $5~\textrm{fm}/c$\hsp . Dots show experimental data~(Lohner et~al. 2013) for (0--40)\%
most central events.
}\label{fig5}
\end{figure}
To study the emission of hard thermal photons we proceed from analytic formulae for chemically
equilibrated QGP suggested by Kapusta et al. (1991). In the lowest-order approximation in strong coupling constant $\alpha_S$ main sources of the real photon production are the $qg$ and $\ov{q}g$ Compton scatterings as well as the $q\ov{q}$ annihilations. Attempts to consider the chemically-nonequilibrium scenario have been already made in (Strickland 1994, K\"ampfer \& Pavlenko 1994, Traxler \& Thoma 1996; Dutta et al. 2002; Gelis et al. 2004). Following their procedure, we include additional suppression factors $\lambda$ and
$\lambda^2$ into the components of photon production corresponding, respectively, to the Compton scattering and $q\overline{q}$ annihilation terms. We have obtained the following expression for the invariant momentum distribution of hard thermal photons (Vovchenko et al. 2015b):
\begin{eqnarray}
&&\mbox{$\frac{\ds dN_\gamma}{\ds d^{\hsp 2}p_{\hsp T} dy}\simeq
\frac{\ds 4\hspm \alpha}{\ds\pi^{\, 3}}\sum\limits_{i=u,d,s}q_i^2{R^{\hsp 2}_{A}}
\int_{\tau_0}^{\tau_f}\tau\hsp d\tau \alpha_s T^2\times$}\label{psm1}\\
&&\hspace*{-2mm}\mbox{$\left\{\lambda^2\left[\ln\left(\frac{a\hspm p_{\hsp T}}
{\alpha_s T}\right) K_0\left(\frac{p_{\hsp T}}{T}\right)+\frac{b\hspm p_{\hsp T}}{T}K_1\left(\frac{p_{\hsp T}}{T}\right)\right]+\lambda\ln\left(\frac{c\hspm p_{\hsp T}}{\alpha_sT}\right)K_0\left(\frac{p_{\hsp T}}{T}\right)\right\}.$}\nonumber
\end{eqnarray}
Here $y$ and $p_{\hsp T}$ are the rapidity and transverse momentum of photons
(it is assumed that \mbox{$p_{\hsp T}\gtrsim T$}\hspm ), and the constants $a\simeq 0.20,~b\simeq 0.99,~c\simeq 0.88$.
Below we disregard the temperature dependence of strong coupling constant, assuming
that $\alpha_s=0.3$.

Figure~\ref{fig5} shows the photon spectra calculated for
the same reaction and same model parameters as above. Again one can see a noticeable
suppression of high $p_T$ photon yields as compared to the equilibrium scenario\hsp\footnote
{
This contradicts the conclusion of Gelis et al. (2004) that chemical nonequilibrium effects
do not modify significantly the photon spectra.
}.
According to Fig.~\ref{fig5} the observed data are better reproduced for
smaller values of $\tau_*$\hsp. Note, that our calculation does not include the contribution of
photons from initial parton-parton collisions. This prediction should be verified in more
realistic calculations, taking into account the transverse flow of deconfined matter.

\section{Evolution of the pure-glue matter}

It is instructive to consider qualitatively the evolution of the idealized pure-glue matter created in relativistic nuclear collisions\hsp\footnote{
Such a case roughly corresponds to small rates of $gg\to q\ov{q}$ reactions. This can be simulated by choosing large $\tau_*$ within the approach developed in Sec.~2. However, we
do not use now the ideal gas approximation for the gluonic matter.
}.
According to the QCD lattice calculations (\mbox{Celik et al.~1983a,} 1983b; Karsch 2002; Bor\hspm s{\'a}nyi et al. 2012, Francis et al. 2015) this quarkless matter should undergo the first order phase transition at the critical temperature $T_c\simeq 270~\textrm{MeV}$. At this temperature the deconfined pure glue matter transforms into the confined state of the pure Yang-Mills theory, namely into a glueball fluid.

\begin{figure}[tp]
\centering
\includegraphics[width=0.9\linewidth]{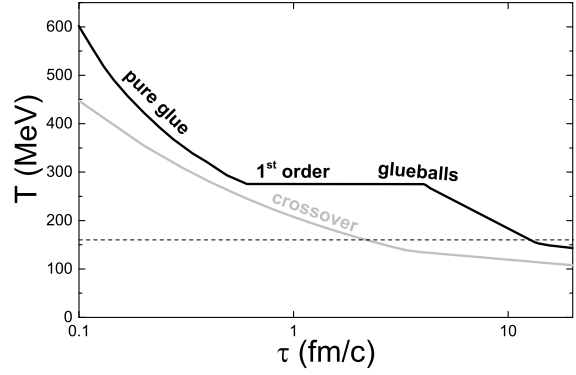}
   \vspace*{4mm}
    \caption[]{Schematic picture of the temperature evolution of a high-energy collision in the pure glue scenario with the Yang-Mills first order phase transition to glueballs.}
\label{fig6}
\end{figure}

Let us assume now that a hot thermalized gluon fluid (with no quarks and antiquarks) is created at the first stage of a relativistic collision. As the system cools and expands, it may reach the mixed phase domain at $T=T_c$ and only after the glue plasma has completely transformed into the glueball fluid, the system cools down further. The system evolution in this pure SU(3) scenario is sketched in Fig.~\ref{fig6}. The possible appearance of super-cooled states and spinodal instabilities, associated with the first-order phase transition, can also be of particular interest.

The heavy glueballs, produced in hadronization of a~pure glue plasma, will evolve (presumably via cascade of two-body decays) into lighter states. Finally the system should decay into hadronic resonances and light hadrons. It was shown within the Frautschi - Hagedorn approach (\mbox{Beitel} et al. 2014)
that the resulting yields of light hadrons and slopes of their spectra agree well
with experimental data on heavy-ion collisions at RHIC and LHC energies.

\begin{figure}[tp]
\centering
\includegraphics[trim=0 2.5cm 0 3cm, clip, width=\linewidth]{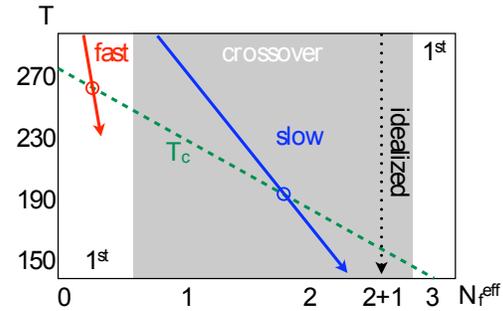}
    \caption[]
    {(Color online)
    Transition temperature of baryon-free QGP versus the effective number of quark flavours.}
\label{fig71}
\end{figure}
In a more realistic scenario one should take into account that some quarks would already be produced before and during the first-order phase transition. Such a scenario could be modeled by introducing the time-dependent effective number of quark degrees of freedom.
With increasing number of quark degrees of freedom the temperature of the phase transition will decrease. At some point the first-order phase transition becomes a smooth crossover. This is schematically shown in Fig.~\ref{fig71}. The qualitative difference between the system evolution in the scenarios with the first-order and crossover transitions is demonstrated in Fig.~\ref{fig6}.
We think that the realization of a particular scenario depends on the energy and size of colliding objects. Thus, future studies of system-size and beam energy dependence of observables in nuclear collisions would be very useful.

\section{Conclusions}
The early stage of high multiplicity pp, pA and AA collision events can
represent a new state of deconfined matter: a~nearly quarkless, pure gluon plasma.
According to the pure Yang - Mills lattice gauge theory, this matter undergoes,
at a~high temperature $T_c \simeq 270$ MeV, the first-order phase transition into
a confined Hagedorn-glueball fluid. Formation of such matter should lead to
suppression of high~$p_T$ photons and dileptons, to reduced baryon to meson ratios,
as well as to enhanced yields of heavy (e.g. charmed) hadrons. We propose to
search for signatures of pure glue states in the LHC/\hsp RHIC and cosmic rays experiments.

\acknowledgements
We would like to thank 
\mbox{L.~McLerran,} Z.~Fodor, S.~Bor\hspm s\'anyi, F.~Karsch, D.~Dietrich, B.~Friman, \mbox{F.~Arleo,}
P.~Giubellino,  C.~Fischer, O.~Philipsen, D.~Vasak, C.~Sturm, H.~Oeschler, K.~Redlich, 
P. Braun-Munzinger, \mbox{S.~Masciocchi,} A.~Andronic, K.-H.~Kampert, 
I.~Kisel, A.M.~Srivastava, \mbox{J.~Aichelin,}  W.~Cassing, E.~Bratkovskaya, \mbox{J.~Harris,} P.~Huovinen,
and H.~Niemi for useful discussions.
This work was supported by the LOEWE Initiative HIC for FAIR of the state of Hesse, by~the Helmholtz Graduate School HIRE for FAIR, by the Frankfurt international graduate school for science, FIGSS at FIAS, by~the Karin and Carlo Giersch Foundation, Frankfurt, by HGF and by BMBF.

\end{document}